\documentclass[review,12pt]{elsarticle}

\usepackage{amssymb,dsfont,amsthm,amsopn}
\usepackage{amsmath}
\usepackage{bm}

\usepackage{amssymb,dsfont,amsthm,amsopn}
\usepackage{amsmath}
\usepackage{bm}

\usepackage{graphicx}
\usepackage[english]{babel}
\usepackage{bm}
\usepackage{setspace}
\usepackage{lineno}
\newtheorem{proposition}{Proposition}

\newtheorem{corollary}{Corollary}

\newcommand{\dsR}{\mathds{R}}
\newcommand{\dsZ}{\mathds{Z}}
\newcommand{\dsD}{\mathds{D}}
\newcommand{\dsE}{\mathds{E}}
\newcommand{\dsA}{\mathds{A}}
\newcommand{\scB}{\mathcal{B}}
\newcommand{\scBd}{{\mathcal{B}^*}}
\newcommand{\scBdw}{{\mathcal{B}^*_w}}
\newcommand{\scP}{\mathcal{P}}
\newcommand{\scO}{\mathcal{O}}
\newcommand{\scC}{\mathcal{C}}
\newcommand{\scI}{\mathcal{I}}
\newcommand{\scU}{\mathcal{U}}
\newcommand{\scG}{\mathcal{G}}
\newcommand{\scS}{\mathcal{S}}

\newcommand{\scE}{\mathcal{E}}

\newcommand{\scDd}{\mathcal{D}^*}

\newcommand{\bv}{\bm{v}}
\newcommand{\bu}{\bm{u}}
\renewcommand{\span}{\operatorname{span}}
\newcommand{\vol}{\operatorname{vol}}
\newcommand{\adj}{\operatorname{adj}}

\begin{document}

\title{A note on sub-orthogonal lattices
}
\author{Jo\~ao Eloir Strapasson}

\address{School of Applied Sciences}

\begin{abstract}
It is shown that, given any $k$-dimensional lattice $\Lambda$, there is a lattice sequence $\Lambda_w$, $w\in \dsZ$, with a sub-orthogonal lattice $\Lambda_o \subset \Lambda$, converging to $\Lambda$ (up less equivalence), also we discuss the conditions for the faster convergence. 
\end{abstract}

\begin{keyword}
	Subortoghonal Lattice \sep Dense Packing \sep Spherical Code \\
	\MSC[2010] 11H31 \sep 11H06 \sep 94A15
\end{keyword}

\maketitle

\section{Introduction}
\label{sec:intro}
A large class of the problems in the coding theory are related to the properties of the lattices, in special, with sublattices generated by orthogonal basis (suborthogonal lattices). Several authors investigated the relationship of the suborthogonal with a spherical codes, and with a q-ary codes, see \cite{carina,sueliperf,rogers,torlayers,toroptimum,vinaycurves}), but, of course, that does not restrict to these problems (\cite{subA2,sequences,arbitrary,complexitylattice}.

In general, the problems in the lattices are concentrated in obtaining certain parameters: as the shortest vector, the packing radius and the packing density, the radius coverage radius and the coverage density.

The points of the lattices are interpreted as elements of a code. Thus finding efficient coding and decoding schemes is essential. There are several schemes in the literature that establish the relationship of the linear codes with the lattices, a good reference is \cite{booksloane}.

This article is organized as follows. In the Section 2, we present the notations, definitions and small properties. In the Section 3, we present a new scheme of obtaining lattices with a sub-orthogonal. In the Section 4, we present a case study for special lattices: $\dsD_n$ and $\dsE_n$ ($n=7,8$) and Leech lattice $\Lambda_{24}$.

\subsection{Background Definitions and Results}
A \textit{lattice} in $\dsR^n$ is an the discrete additive subgroup of $\dsR^n$, $\Lambda$, which has a \textit{generator matrix} with full rank, $k\times n$, $\scB$, e.g, $\bv \in \Lambda \leftrightarrow \bv=\bu^t \scB$ ($\bu \in \dsZ^k$, $k$ is said rank of $\Lambda$. The \textit{determinant} of a lattice is $\det(\Lambda)=\det(\scG)$, there $\scG=\scB \scB^t$ is a \textit{Gram matrix} of the lattice $\Lambda$ and the \textit{volume} of the lattice is $\sqrt{\det(\Lambda)}$ (volume of the parallelotope generate for rows of $\scB$). The \textit{minimum norm} of the Lattice $\Lambda$, $\rho(\Lambda)$, is $\min\{\|\bv\|;\bv\in\Lambda \text{ and } \bv \neq \bm 0\}$ and \textit{center density packing} of $\Lambda$ is $\delta_\Lambda=\dfrac{\rho(\Lambda)^n}{2^n \vol(\Lambda)}$. Two lattices $\Lambda_1$ and $\Lambda_2$, with generator matrices $\scB_1$ and $\scB_2$ are \textit{equivalence} if, only if $\scB_1=c\, \scU \scB_2 \scO$, there $c\in\dsR$, $\scU$ is \textit{uni-modular matrix} (integer, $k\times k$ matrix with $\det(\scU)=\pm 1$) and $\scO$ is the orthogonal, $n\times n$ matrix ($\scO \scO^t= \scI_n$, $\scI_n$ identity matrix $n\times n$). \textit{Dual lattice} of $\Lambda$ is a lattice, $\Lambda^*$, obtained for all vectors $\bu \in \span(\scB)$ (there $\span(\scB$) is a vector space generated by the rows of $\scB$) with that $\bu\cdot \bv\in \dsZ, \forall \bv \in \Lambda$, the generator matrix of $\Lambda^*$ is $\scBd=(\scB \scB^t)^{-1} \scB$, in particular, $\scBd=\scB^{-t}$ if $n=k$. A sub-lattice, $\Lambda'$, is a subset of $\Lambda$ which is also lattice, if $\Lambda'$ has generator matrix is formed by the orthogonal row vectors we will say that it is a sub-orthogonal lattice.

Since the lattice is a group, remember that the quotiente of a lattice $\Lambda$ by a sublattice $\Lambda'$, $\frac{\Lambda}{\Lambda'}$, is as a finite abelian group with $M$ elements, where $M$ is the ratio of the volume of $\Lambda'$ by the volume of $\Lambda$, e.g., $M=\frac{\vol(\Lambda')}{\vol(\Lambda)}$. The $M$ elements of the lattice $\Lambda$, can be seen as an orbit of vector in the $k$-dimensional torus $\frac{\Lambda}{\Lambda'}$. This essentially establishes the relationship with a central spherical class codes, as well as a class of the linear codes track construction ``A" and similar constructions, see more details in \cite{booksloane}.

\section{Suborthogonal sequences}
Consider the lattice $\Lambda \subset \dsR^n$, of rank $n$, contain an orthogonal sub-lattice, $\Lambda_o \subset \Lambda$, such that $\Lambda_o$ is equivalence to $\dsZ^n$, e.g., the generator matrix of the $\Lambda_o$ is $c\, \scO$, with $\scO\scO^t =\scI_n$. Let $\scB$ and $\scBd=\scB^{-t}$ the generator matrices of the $\Lambda$ and $\Lambda^*$ (respectively). Assuming that $\scBd$ has integer entries, Then the lattice, $\Lambda$, with generator matrix $\scB=\adj(\scBd)=\det(\scBd) \scBd^{-t}$ has a sub-orthogonal lattice, $\Lambda_o$, with generator matrix $\scBd \scB =\det(\scBd) \scI_n$. The ratio of the volume measured quantities points and in this case it is $\frac{\vol(\Lambda_o)}{\vol(\Lambda)}=\frac{\det(\scBd \scB)}{\det(\scB)}=\det(\scBd)$. 

We want to build codes with a large number of the points. And we observe that as we want to increase the number of points we must increase the determinant of the matrix of the dual lattice.

\begin{proposition}\label{Construction}
	 Let $\Lambda$ be the $n$-dimensional lattice and $\Lambda^*$ be its dual, with generator matrices $\scB$ and $\scBd$ (respectively). Assuming that $\scBd$ has integer entries. Define $\Lambda_w^*$ with generator matrix $\scBdw=w \scBd+\scP$ ($w$ is integer and $\scP$ is an integer matrix any). Then the lattices $\Lambda^*_w$ and $\Lambda_w$ with generator matrices $\scBdw$ and $\scB_w=\adj(\scBdw)$ (respectively) to satisfy $\frac{1}{w} \Lambda^*_w \longrightarrow \Lambda^*$ ($w\to\infty$) and by continuity of the matrix inversion process $\frac{1}{\det(\frac{1}{w} \scBdw)} \Lambda_w\longrightarrow \Lambda$ (${w\to\infty}$).
\end{proposition}

\begin{proof}
	The proof is trivial by the fact that the convergence of each entry of the generator matrix, and the convergence of the generating matrix defines convergence groups. Recalling that the cardinality of the points in the quotient, $\Lambda_w^*$ is a polynomial, specifically $M(w)=\det(\Lambda^*_w).$
\end{proof}

\begin{corollary}\label{coroConstruction}
	Let $\Lambda$ be the $n$-dimensional lattice and $\Lambda^*$ be its dual, with a generator matrices $\scB$ and $\scBd$ (respectively). Define $\Lambda_w^*$ with generator matrix $\scBdw=w \scBd+\scP$ ($\scP=\lfloor w \scBd\rceil - w \scBd$), in other words, $\scBdw=\lfloor w \scBd\rceil$ (rounded entries). Then the lattices $\Lambda^*_w$ and $\Lambda_w$ with generator matrices $\scB_w$ and $\scB_w=\adj(\scBdw)$ (respectively) to satisfy $\frac{1}{w} \Lambda^*_w \longrightarrow \Lambda^*$ (${w\to\infty}$) and by continuity of the matrix inversion process $\frac{1}{\det(\frac{1}{w} \scBdw)} \Lambda_w\longrightarrow \Lambda$ (${w\to\infty}$).
\end{corollary}

The corollary allows to extend the use of the proposition for whose lattices dual have not, up less equivalence, integer generator matrix. 

In the following propositions we establish the speed of the convergence:	

\begin{proposition}[faster dual convergence]\label{fasterDualConvergence}
Let $\Lambda^*_w$ and $\Lambda_w$ as in Proposition \ref{Construction}. Then faster convergence is obtained minimizing inputs $\scP\scP^t$, considering $\scP=\scS$ $\scB$ ($\scS$ is antisymmetric matrix $n\times n$, whose parameters will be minimized). Of course, if $\scP$ is identically zero is the best convergence, because there is no error.
\end{proposition} 
\begin{proof}
	A better sequence is that in which the vectors is closest sizes and angles of the desired lattice. And so, we must analyse the sequence formed by Gram matrix $1/w^2 \scG^*_w=(1/w \scB_w)(1/w \scB_w)^t$. Therefore
	\begin{eqnarray*}
		1/w^2 \scG^*_w & = & (1/w \scBdw)(1/w \scBdw)^t\\
		& = & (\scBd+1/w \scP)(\scBd+1/w \scP)^t\\
		& = & (\scBd+1/w \scP)(\scBd^t+1/w \scP^t)\\
		& = & \scBd \scBd^t+1/w (\scP \scBd^t+\scBd \scP^t)+1/w^2 \scP \scP^t
	\end{eqnarray*}
	
	From now on we will say that convergence is linear if $\scP \scBd^t+ \scBd \scP^t=\alpha \scBd$ $\scBd^t$ and quadratic if $\scP \scBd^t+ \scBd \scP^t=\alpha \scBd \scBd^t$ and constant if $\scP=0$ ($\scP$ identically zero), unless a change in $w$ variable, we can assume $\alpha=0$, in these conditions:
	\begin{eqnarray*}
		\scP \scBd^t+\scBd \scP^t=0 \Rightarrow \scP \scBd^t=-\scBd \scP^t=-(\scP \scBd^t)^t \textbf{(skew-symmetric)}.
	\end{eqnarray*}
	Then $\scP =\scS \scBd^{-t}=\scS \scB$, there any antisymmetric $\scS$ with $\scP =\scS \scB$ integer matrix. In the case, of convergence is quadratic, the convergence coefficient also depends on the inputs $\scP\scP^t$, which must be minimized.
\end{proof}

\begin{proposition}[faster convergence]\label{fasterConvergence}
		Let $\Lambda^*_w$ and $\Lambda_w$ as in Proposition \ref{Construction}.  Then faster convergence is obtained minimizing inputs $\scB \scP^t \scG \scP \scB^t$, considering $\scP=\scBd \scS$ ($\scS$ is antisymmetric matrix $n\times n$, whose parameters will be minimized). Of course, if $\scP$ is identically zero is the best convergence, because there is no error.
	\end{proposition}

\begin{proof}
	 We recall that the inverse of a sum of matrix with identity matrix can be calculated by Neumann series (\cite{bookmatrices}) $(A+I)^{-1}=\sum_{n = 0}^\infty	(-A)^n)$, so the dual generator matrix of lattice sequence is:
	 \begin{eqnarray*}
		 \scBdw & = & w \scBd+\scP\\
		 & = & w \scBd (\scI_n \scBd^{-1} \scP)\\
		 & = & w \scBd (\scI_n+1/w\scB^t \scP),
	 \end{eqnarray*}
	 and the inverse transpose is:
	 \begin{eqnarray*}
	 \scB_w & = & \adj(\scBdw)\\
	 & = & \det(\scBdw)\scBdw^{-t} (\beta:=\det(\scBdw))\\
	 & = & \beta (w \scBd (\scI_n+1/w \scB^t \scP))^{-t}\\
	 & = & \beta/w^n \scBd^{-t}(\scI_n+1/w \scB^t \scP)^{-t}\\
	 & = & \beta/w^n \scB (\scI_n+1/w \scP^t \scB)^{-1}\\
	 & = & \beta/w^n \scB (\scI_n+1/w \scP^t \scB)^{-1}\\
	 & = & \beta/w^n \scB (\scI_n-1/w \scP^t \scB+(1/w \scP^t \scB)^2-1/w \scP^t \scB)^3+\cdots)\\
	 & \approx & \beta/w^n \scB (\scI_n-1/w \scP^t \scB).
	 \end{eqnarray*}
	 
	 Therefore, the gram matrix approximated is:
	 \begin{eqnarray*}
	 	\scG_w & \approx & (\beta/w)^2 \scB(\scI_n-1/w \scP^t \scB)(\scI_n-1/w \scP^t \scB)^t \scB^t \\
	 	& = & (\beta/w)^2 \scB(\scI_n-1/w \scP^t \scB)(\scI_n-1/w \scB^t \scP)\scB^t\\
	 	& = & (\beta/w)^2 \scB(\scI_n-1/w (\scP^t \scB+\scB^t \scP)+1/w^2 \scP^t \scG \scP)\scB^t\\
	 	& = & (\beta/w)^2 (\scG-1/w \scB(\scP^t \scB+\scB^t \scP)\scB^t+1/w^2 \scB \scP^t \scG \scP \scB^t).
	 \end{eqnarray*}

	 It is desirable that:
	 \begin{eqnarray*}
	 	\scP^t \scB+\scB^t \scP=0 \Rightarrow \scB^t \scP=-\scP^t\scB=-(\scB^t \scP)^t=\scS  (\textbf{skew-symmetric}),
	\end{eqnarray*}
	from which it follows that $\scP=\scBd \scS$, there any antisymmetric $\scS$ with $\scP=\scBd \scS$ integer matrix. In the case, of convergence is quadratic, the convergence coefficient also depends on the inputs $\scB \scP^t \scG \scP \scB^t$, which must be minimized.
\end{proof}

The structure of the group obtained by the quotient of the lattice sequence, $\Lambda_w$, by their respective an orthogonal sublattice can be determined and extended, applying the Theorem 2.4.13 in \cite{bookcohen}. 

In particular, $\scBd$ is lower triangular matrix and $\scP=\scC_n=(c_{i,j})$ (cyclic perturbation), where $c_{i,j}=1$ if $j=i+1$ and $c_{i,j}=0$ otherwise, the quotient is cyclic group although convergence is not nearly quadratic.

Lattices of rank $n$ that, up less equivalence, are sublattices the integer lattice $\dsZ^n$, play an interesting role with regard to the convergence as discussed below with case study, next section.

\section{Case study}
In this section, we present a construction applied to the special cases: $\dsD_n $, $\dsE_n (n = 7,8) $ and $\Lambda24 $ and we show the best perturbations found. Although the null perturbation is optimal, it is never associated with cyclic quotient groups, it does not offer the best solution in terms of the spherical codes, as we shall see below. Being cyclical and optimal is unlikely. In addition, the results presented here complement results obtained in \cite{rogers} (not considering the initial vector problem) and extend and simplify the results obtained in \cite{carina}.

\subsection{The root lattice $\dsD_n$ ($n\geq3$)}
We consider the generate matrix of $\dsD_n^*$ as $\scDd_n$ and good perturbation is $\scP_n$: 

\begin{equation}\label{eq:dn}
\scDd_n=\begin{bmatrix}
2 & 0 & \cdots & 0 & 0 \\
0 & 2 & 0 & \cdots & 0 \\
\vdots & \vdots & \ddots & \vdots & \vdots \\
0 & 0 & \cdots & 2 & 0 \\
1 & 1 & \cdots & 1 & 1
\end{bmatrix} \mbox{and }
\scP_n=\begin{bmatrix}
0 & 1 & 0 & \cdots & 0 & 0 & 1 \\
-1 & 0 & 1 & \cdots & 0 & 0 & 0 \\
0 & -1 & 0 & \ddots & 0 & 0 & 0 \\
\vdots & \vdots & \ddots & \ddots & \ddots & \vdots & \vdots \\
0 & 0 & 0 & \ddots & 0 & 1 & 0 \\
0 & 0 & 0 & \cdots & -1 & 0 & 1 \\
-1 & 0 & 0 & \cdots & 0 & 0 & 1
\end{bmatrix}.
\end{equation}

The good perturbation is $\scP_n$ in \eqref{eq:dn}, this case the quotient is cyclic case odd $n$, the performance is illustrated in Tables \ref{table1} and \ref{table2}.

\begin{table}[!h]
$$	
\arraycolsep=1pt\def\arraystretch{1}
\begin{array}{|ccc|ccc|ccc|}
 \hline
 M(w) & \delta(\Lambda_w) & \text{Group} & M(w) & \delta(\Lambda_w) & \text{Group} & M(w) & \delta(\Lambda_w) & \text{Group}\\ \hline
 4 & 0.176777 & \dsZ_2\oplus \dsZ_2 & 7 & 0.133631 & \dsZ_7 & 3 & 0.0721688 & \dsZ_3 \\
 32 & 0.176777 & \dsZ_2\oplus \dsZ_4\oplus \dsZ_4 & 38 & 0.162221 & \dsZ_{38} & 26 & 0.0969021 & \dsZ_{26} \\
 108 & 0.176777 & \dsZ_3\oplus \dsZ_6\oplus \dsZ_6 & 117 & 0.169842 & \dsZ_{117} & 93 & 0.116923 & \dsZ_{93} \\
 256 & 0.176777 & \dsZ_4\oplus \dsZ_8\oplus \dsZ_8 & 268 & 0.172774 & \dsZ_{268} & 228 & 0.129349 & \dsZ_{228} \\
 500 & 0.176777 & \dsZ_5\oplus \dsZ_{10}\oplus \dsZ_{10} & 515 & 0.174183 & \dsZ_{515} & 455 & 0.137602 & \dsZ_{455} \\
 864 & 0.176777 & \dsZ_6\oplus \dsZ_{12}\oplus \dsZ_{12} & 882 & 0.174964 & \dsZ_{882} & 798 & 0.143442 & \dsZ_{798} \\
 1372 & 0.176777 & \dsZ_7\oplus \dsZ_{14}\oplus \dsZ_{14} & 1393 & 0.175439 & \dsZ_{1393} & 1281 & 0.147780 & \dsZ_{1281} \\
 2048 & 0.176777 & \dsZ_8\oplus \dsZ_{16}\oplus \dsZ_{16} & 2072 & 0.175750 & \dsZ_{2072} & 1928 & 0.151126 & \dsZ_{1928} \\
 2916 & 0.176777 & \dsZ_9\oplus \dsZ_{18}\oplus \dsZ_{18} & 2943 & 0.175964 & \dsZ_{2943} & 2763 & 0.153783 & \dsZ_{2763} \\
 4000 & 0.176777 & \dsZ_{10}\oplus \dsZ_{20}\oplus \dsZ_{20} & 4030 & 0.176117 & \dsZ_{4030} & 3810 & 0.155943 & \dsZ_{3810} \\ \hline
\end{array}$$
\caption{Show performance in 3-dimensional case, for perturbations $\bm 0_n$, $\scP_n$ and $\scC_n$ respectively.}
	\label{table1}
\end{table}

\begin{table}[!h]
	\arraycolsep=1pt\def\arraystretch{1}
$$\begin{array}{|cc|cc|cc|cc|}\hline
	\frac{\delta(\Lambda_w)}{\delta(\dsD_3)} & \text{Group} & \frac{\delta(\Lambda_w)}{\delta(\dsD_4)} & \text{Group} & \frac{\delta(\Lambda_w)}{\delta(\dsD_5)} & \text{Group} & \frac{\delta(\Lambda_w)}{\delta(\dsD_6)} & \text{Group} \\ \hline
	0.7559 & \dsZ_7 & 1. & \dsZ_3\oplus \dsZ_6 & 0.6718 & \dsZ_{41} & 0.675 & \dsZ_{10}\oplus \dsZ_{10} \\
	0.9177 & \dsZ_{38} & 1. & \dsZ_9\oplus \dsZ_{18} & 0.8732 & \dsZ_{682} & 0.8576 & \dsZ_{17}\oplus \dsZ_{170} \\
	0.9608 & \dsZ_{117} & 1. & \dsZ_{19}\oplus \dsZ_{38} & 0.9371 & \dsZ_{4443} & 0.9269 & \dsZ_{74}\oplus \dsZ_{370} \\
	0.9774 & \dsZ_{268} & 1. & \dsZ_{33}\oplus \dsZ_{66} & 0.9631 & \dsZ_{17684} & 0.9565 & \dsZ_{65}\oplus \dsZ_{2210} \\
	0.9853 & \dsZ_{515} & 1. & \dsZ_{51}\oplus \dsZ_{102} & 0.9759 & \dsZ_{52525} & 0.9714 & \dsZ_{202}\oplus \dsZ_{2626} \\
	0.9897 & \dsZ_{882} & 1. & \dsZ_{73}\oplus \dsZ_{146} & 0.9831 & \dsZ_{128766} & 0.9799 & \dsZ_{145}\oplus \dsZ_{10730} \\
	0.9924 & \dsZ_{1393} & 1. & \dsZ_{99}\oplus \dsZ_{198} & 0.9875 & \dsZ_{275807} & 0.9851 & \dsZ_{394}\oplus \dsZ_{9850} \\
	0.9942 & \dsZ_{2072} & 1. & \dsZ_{129}\oplus \dsZ_{258} & 0.9904 & \dsZ_{534568} & 0.9885 & \dsZ_{257}\oplus \dsZ_{33410} \\
	0.9954 & \dsZ_{2943} & 1. & \dsZ_{163}\oplus \dsZ_{326} & 0.9924 & \dsZ_{959409} & 0.9909 & \dsZ_{650}\oplus \dsZ_{26650} \\
	0.9963 & \dsZ_{4030} & 1. & \dsZ_{201}\oplus \dsZ_{402} & 0.9938 & \dsZ_{1620050} & 0.9926 & \dsZ_{401}\oplus \dsZ_{81002} \\ \hline
\end{array}$$
	\caption{Show performance in 3 to 6-dimensional case, for perturbations $\scP_n$.}
	\label{table2}
\end{table}

\subsection{The root lattice $\dsE_n$}
Up less equivalence, assuming that $\dsE_7^*$, $\dsE_{8,1}^*$ and $\dsE_{8,2}^*$ are generated by matrices $\scE_7^*$, $\scE_{8,1}^*$ and $\scE_{8,2}^*$ and the good perturbation $\scP_7$, $\scP_{8,1}$ and $\scP_{8,2}$. 

$$
\scE_7^*=\arraycolsep=2.5pt
\begin{bmatrix}
 1 & 0 & 0 & 0 & -1 & 0 & 1 \\
 0 & 1 & 0 & 0 & -1 & -1 & 1 \\
 0 & 0 & 1 & 0 & -1 & -1 & 0 \\
 0 & 0 & 0 & 1 & 0 & -1 & -1 \\
 0 & 0 & 0 & 0 & 2 & 0 & -2 \\
 0 & 0 & 0 & 0 & 0 & 2 & 0 \\
 0 & 0 & 0 & 0 & 0 & 0 & 2 \\
\end{bmatrix}, 
\scP_7=\arraycolsep=2.5pt
\begin{bmatrix}
 0 & 1 & 0 & 0 & 0 & 0 & 0 \\
 -1 & 0 & 1 & 0 & 0 & 0 & 0 \\
 0 & -1 & 0 & 0 & 0 & 0 & 0 \\
 -1 & -1 & -1 & 0 & -1 & 0 & 0 \\
 0 & 0 & -1 & 1 & -1 & 0 & -1 \\
 0 & 0 & 1 & 1 & 1 & 0 & 1 \\
 1 & 0 & 0 & -1 & 1 & -1 & 0 \\
\end{bmatrix}
$$

$$
\scE_{8,1}^*=\arraycolsep=2.5pt
\begin{bmatrix}
 1 & 0 & 0 & 0 & -1 & 0 & 1 & 1 \\
 0 & 1 & 0 & 0 & -1 & -1 & 1 & 0 \\
 0 & 0 & 1 & 0 & -1 & -1 & 0 & 1 \\
 0 & 0 & 0 & 1 & 0 & -1 & -1 & -1 \\
 0 & 0 & 0 & 0 & 2 & 0 & -2 & -2 \\
 0 & 0 & 0 & 0 & 0 & 2 & 0 & 0 \\
 0 & 0 & 0 & 0 & 0 & 0 & 2 & 0 \\
 0 & 0 & 0 & 0 & 0 & 0 & 0 & 2 \\
\end{bmatrix},
\scP_{8,1}=
\arraycolsep=2.5pt
\begin{bmatrix}
 0 & 1 & 1 & 0 & 0 & 0 & 0 & 0 \\
 -1 & 0 & 0 & 0 & 0 & 0 & 0 & 0 \\
 0 & 1 & 0 & -1 & 0 & 0 & -1 & 0 \\
 0 & 0 & 0 & 0 & 0 & 0 & 0 & 0 \\
 0 & -1 & -1 & 0 & 0 & 0 & 1 & -1 \\
 0 & 0 & 1 & 1 & 1 & 0 & 1 & 0 \\
 -1 & -1 & 0 & 0 & 0 & -1 & 0 & 1 \\
 1 & 1 & 0 & -1 & 0 & 0 & -1 & 0 \\
\end{bmatrix}
$$

$$
\scE_{8,2}^*=
\arraycolsep=2.5pt
\begin{bmatrix}
 1 & 1 & 1 & 1 & 1 & 1 & 1 & -7 \\
 -1 & 1 & 1 & 1 & 1 & 1 & 1 & -5 \\
 0 & 0 & 2 & 2 & 2 & 2 & 2 & -10 \\
 0 & 0 & 0 & 2 & 2 & 2 & 2 & -8 \\
 0 & 0 & 0 & 0 & 2 & 2 & 2 & -6 \\
 0 & 0 & 0 & 0 & 0 & 2 & 2 & -4 \\
 0 & 0 & 0 & 0 & 0 & 0 & 2 & -2 \\
 0 & 0 & 0 & 0 & 0 & 0 & 0 & 4 \\
\end{bmatrix},
\scP_{8,2}=
\arraycolsep=2.5pt
\begin{bmatrix}
 -1 & 0 & 0 & 0 & 0 & 0 & 1 & 0 \\
 -1 & -1 & 0 & 0 & 0 & 0 & 0 & 0 \\
 -1 & -1 & -1 & 0 & 1 & 0 & 0 & 0 \\
 -1 & 0 & -1 & -1 & 0 & 1 & 0 & 0 \\
 -1 & 0 & 0 & -1 & 0 & 0 & 0 & 0 \\
 -1 & -1 & 1 & -1 & 0 & -1 & 1 & 0 \\
 -1 & -1 & 0 & 0 & 1 & -1 & 0 & 0 \\
 0 & 0 & 0 & 0 & 0 & 0 & 0 & 0 \\
\end{bmatrix}
.
$$

The performance is illustrated in Table \ref{table3} (note that the density ratio is deployed close and the amount of the associated points are: 1.664.641.200 points for dual lattice $10\, \dsE_{8,1}^*+\scP_{8,1}$ and 11.430.630.576 for dual lattice $9\, \dsE_{8,2}^*+\scP_{8,2}$ (very more points in the second case). The Table \ref{table4}, illustrates the performance applied in spherical codes, details in \cite{rogers}, the non-null perturbation is better in the case of $\dsE_{8,1}$ representation, moreover, point out that the performance is similar to the second representation with null perturbation.

\begin{table}[!h]
	\arraycolsep=1pt\def\arraystretch{1}
	$$\begin{array}{|cc|cc|cc|}\hline
\frac{\delta(\Lambda_w)}{\delta(\dsE_7)} & \text{Group} & \frac{\delta(\Lambda_w)}{\delta(\dsE_{8,1})} & \text{Group} & \frac{\delta(\Lambda_w)}{\delta(\dsE_{8,2})} & \text{Group} \\ \hline
 0.2346 & \dsZ_2\oplus \dsZ_{68} & 0.1204 & \dsZ_2\oplus \dsZ_{78} & 0.2706 & \dsZ_2\oplus \dsZ_2\oplus \dsZ_{364} \\
 0.4161 & \dsZ_2\oplus \dsZ_{1552} & 0.4022 & \dsZ_4\oplus \dsZ_{2316} & 0.5065 & \dsZ_2\oplus \dsZ_4\oplus \dsZ_{15128} \\
 0.5966 & \dsZ_2\oplus \dsZ_{15468} & 0.622 & \dsZ_6\oplus \dsZ_{26154} & 0.6918 & \dsZ_2\oplus \dsZ_6\oplus \dsZ_{189252} \\
 0.7208 & \dsZ_2\oplus \dsZ_{92192} & 0.7521 & \dsZ_8\oplus \dsZ_{165912} & 0.7993 & \dsZ_2\oplus \dsZ_8\oplus \dsZ_{1251376} \\
 0.8005 & \dsZ_2\oplus \dsZ_{391540} & 0.8284 & \dsZ_{10}\oplus \dsZ_{729030} & 0.8616 & \dsZ_2\oplus \dsZ_{10}\oplus \dsZ_{5612060} \\
 0.8522 & \dsZ_2\oplus \dsZ_{1313328} & 0.8754 & \dsZ_{12}\oplus \dsZ_{2495268} & 0.8997 & \dsZ_2\oplus \dsZ_{12}\oplus \dsZ_{19429704} \\
 0.8869 & \dsZ_2\oplus \dsZ_{3708572} & 0.9059 & \dsZ_{14}\oplus \dsZ_{7137186} & 0.9243 & \dsZ_2\oplus \dsZ_{14}\oplus \dsZ_{55966708} \\
 0.911 & \dsZ_2\oplus \dsZ_{9191488} & 0.9266 & \dsZ_{16}\oplus \dsZ_{17842224} & 0.9411 & \dsZ_2\oplus \dsZ_{16}\oplus \dsZ_{140558432} \\
 0.9284 & \dsZ_2\oplus \dsZ_{20572452} & 0.9412 & \dsZ_{18}\oplus \dsZ_{40176702} & \bm{0.9529} & \dsZ_2\oplus \dsZ_{18}\oplus \dsZ_{317517516} \\
 0.9412 & \dsZ_2\oplus \dsZ_{42432080} & \bm{0.952} & \dsZ_{20}\oplus \dsZ_{83232060} & 0.9615 & \dsZ_2\oplus \dsZ_{20}\oplus \dsZ_{659296120} \\ \hline
\end{array}$$
	\caption{Show performance in 7 to 8-dimensional case, for representations $\dsE_{7}^*$, $\dsE_{8,1}^*$ $\dsE_{8,2}^*$ and perturbations $\scP_7$, $\scP_{8,1}$ and $\scP_{8,2}$.}
	\label{table3}
\end{table}

\begin{table}[!h]
\centering	
\begin{tabular}{|ccc|}
\hline
 & $\dsE_{8,1}$ & $\dsE_{8,2}$ \\ \hline \hline
 $\bm 0$ &  $
\begin{array}{|cc|}
\text{Distance} & M-\text{Points} \\ \hline
 0.707107 & 4096 \\
 0.707107 & 104976 \\
 0.500000 & 1048576 \\
 0.415627 & 6250000 \\
 0.366025 & 26873856 \\
 0.306802 & 92236816 \\
 0.270598 & 268435456 \\
 0.241845 & 688747536 \\
 0.218508 & 1600000000 \\
\end{array}$ & $\begin{array}{|ccc|}
\text{Distance} & M-\text{Points} \\ \hline
 0.707107 & 65536 \\
 0.500000 & 1679616 \\
 0.382683 & 16777216 \\
 \textbf{0.309017} & \textbf{100000000} \\
 0.258819 & 429981696 \\
 0.222521 & 1475789056 \\
 0.195090 & 4294967296 \\
 0.173648 & 11019960576 \\
 0.156434 & 25600000000 \\
\end{array}$ \\ \hline
 $\scP_{8,i}$ & $
\begin{array}{|cc|}
\text{Distance} & M-\text{Points} \\ \hline
 0.839849 & 9264 \\
 0.641669 & 156924 \\
 0.509472 & 1327296 \\
 0.419589 & 7290300 \\
 0.355527 & 29943216 \\
 \textbf{0.307914} & \textbf{99920604} \\
 0.271283 & 285475584 \\
 0.242296 & 723180636 \\
 0.218821 & 1664641200 \\
\end{array}$ & $
\begin{array}{|cc|}
\text{Distance} & M-\text{Points} \\ \hline
 0.639702 & 121024 \\
 0.468092 & 2271024 \\
 0.366403 & 20022016 \\
 0.299852 & 112241200 \\
 0.253223 & 466312896 \\
 0.218878 & 1567067824 \\
 0.192596 & 4497869824 \\
 0.171869 & 11430630576 \\
 0.155124 & 26371844800 \\
\end{array}$ \\ \hline
\end{tabular}
	\caption{Show spherical code performance 8-dimensional case, for different representations.}
	\label{table4}
\end{table}

\subsection{The Leech lattices $\Lambda_{24}$}
The laminate lattice is generally dense in their respective dimensions in special dimensions, $n=9,15,16,19,20,21,24$ admit integer representation, un less equivalence, and in these cases can analyse the fast convergence, consider $n=24$ the matrix generator of Leech Lattice, unless equivalence is:

$$\mathfrak{L}_{24,1}=
\left[\begin{smallmatrix}
 4 & 0 & 0 & 0 & 0 & 0 & 0 & 0 & 0 & 0 & 0 & 0 & 0 & 0 & 0 & 0 & 0 & 0 & 0 & 0 & 0 & 0 & 0 & 0 \\
 0 & 4 & 0 & 0 & 0 & 0 & 0 & 0 & 0 & 0 & 0 & 0 & 0 & 0 & 0 & 0 & 0 & 0 & 0 & 0 & 0 & 0 & 0 & 0 \\
 0 & 0 & 4 & 0 & 0 & 0 & 0 & 0 & 0 & 0 & 0 & 0 & 0 & 0 & 0 & 0 & 0 & 0 & 0 & 0 & 0 & 0 & 0 & 0 \\
 0 & 0 & 0 & 4 & 0 & 0 & 0 & 0 & 0 & 0 & 0 & 0 & 0 & 0 & 0 & 0 & 0 & 0 & 0 & 0 & 0 & 0 & 0 & 0 \\
 2 & 2 & 2 & 0 & 2 & 0 & 0 & 0 & 0 & 0 & 0 & 0 & 0 & 0 & 0 & 0 & 0 & 0 & 0 & 0 & 0 & 0 & 0 & 0 \\
 0 & 2 & 2 & 2 & 0 & 2 & 0 & 0 & 0 & 0 & 0 & 0 & 0 & 0 & 0 & 0 & 0 & 0 & 0 & 0 & 0 & 0 & 0 & 0 \\
 0 & 0 & 2 & 2 & 2 & 0 & 2 & 0 & 0 & 0 & 0 & 0 & 0 & 0 & 0 & 0 & 0 & 0 & 0 & 0 & 0 & 0 & 0 & 0 \\
 0 & 2 & 0 & 2 & 2 & 0 & 0 & 2 & 0 & 0 & 0 & 0 & 0 & 0 & 0 & 0 & 0 & 0 & 0 & 0 & 0 & 0 & 0 & 0 \\
 0 & 0 & 0 & 0 & 0 & 0 & 0 & 0 & 4 & 0 & 0 & 0 & 0 & 0 & 0 & 0 & 0 & 0 & 0 & 0 & 0 & 0 & 0 & 0 \\
 0 & 0 & 0 & 0 & 0 & 0 & 0 & 0 & 0 & 4 & 0 & 0 & 0 & 0 & 0 & 0 & 0 & 0 & 0 & 0 & 0 & 0 & 0 & 0 \\
 0 & 0 & 0 & 0 & 0 & 0 & 0 & 0 & 0 & 0 & 4 & 0 & 0 & 0 & 0 & 0 & 0 & 0 & 0 & 0 & 0 & 0 & 0 & 0 \\
 0 & 0 & 0 & 0 & 0 & 0 & 0 & 0 & 0 & 0 & 0 & 4 & 0 & 0 & 0 & 0 & 0 & 0 & 0 & 0 & 0 & 0 & 0 & 0 \\
 -1 & 1 & 1 & 2 & 1 & 0 & 0 & 0 & -1 & 1 & 1 & 2 & 1 & 0 & 0 & 0 & 0 & 0 & 0 & 0 & 0 & 0 & 0 & 0 \\
 2 & 1 & -1 & 1 & 0 & 1 & 0 & 0 & 2 & 1 & -1 & 1 & 0 & 1 & 0 & 0 & 0 & 0 & 0 & 0 & 0 & 0 & 0 & 0 \\
 1 & 1 & -2 & -1 & 0 & 0 & 1 & 0 & 1 & 1 & 2 & -1 & 0 & 0 & 1 & 0 & 0 & 0 & 0 & 0 & 0 & 0 & 0 & 0 \\
 1 & 2 & 1 & 1 & 0 & 0 & 0 & 1 & 1 & 2 & 1 & 1 & 0 & 0 & 0 & 1 & 0 & 0 & 0 & 0 & 0 & 0 & 0 & 0 \\
 -1 & 1 & 2 & -1 & 0 & 0 & 1 & 0 & 2 & 0 & 0 & 0 & 0 & 0 & 0 & 0 & 2 & 0 & 0 & 0 & 0 & 0 & 0 & 0 \\
 -1 & -1 & -1 & -1 & 1 & 1 & 1 & 1 & 0 & 2 & 0 & 0 & 0 & 0 & 0 & 0 & 0 & 2 & 0 & 0 & 0 & 0 & 0 & 0 \\
 2 & -1 & 1 & 0 & 0 & 0 & 1 & 1 & 0 & 0 & 2 & 0 & 0 & 0 & 0 & 0 & 0 & 0 & 2 & 0 & 0 & 0 & 0 & 0 \\
 1 & 1 & 2 & 1 & 0 & 0 & 1 & 0 & 0 & 0 & 0 & 2 & 0 & 0 & 0 & 0 & 0 & 0 & 0 & 2 & 0 & 0 & 0 & 0 \\
 1 & 1 & 1 & 0 & 1 & 0 & 0 & 0 & 2 & 0 & 0 & 2 & 0 & 0 & 0 & 0 & 1 & 1 & 1 & 0 & 1 & 0 & 0 & 0 \\
 1 & -1 & 2 & 2 & 0 & 1 & 0 & 1 & 2 & 0 & 2 & 0 & 0 & 0 & 0 & 0 & 0 & 1 & 1 & 1 & 0 & 1 & 0 & 0 \\
 1 & 0 & 1 & 0 & 0 & -1 & -1 & 0 & 0 & -2 & 2 & 0 & 0 & 0 & 0 & 0 & 0 & 0 & 1 & 1 & 1 & 0 & 1 & 0 \\
 2 & 2 & 0 & -1 & 2 & 1 & -1 & -1 & 0 & 2 & -2 & 2 & 0 & 0 & 0 & 0 & 0 & 1 & 0 & 1 & 1 & 0 & 0 & 1 \\
\end{smallmatrix}
\right], \mbox{ or }$$ 

$$
\mathfrak{L}_{24,2}=
\left[\begin{smallmatrix}
	4 & 4 & 0 & 0 & 0 & 0 & 0 & 0 & 0 & 0 & 0 & 0 & 0 & 0 & 0 & 0 & 0 & 0 & 0 & 0 & 0 & 0 & 0 & 0 \\
	-4 & 4 & 0 & 0 & 0 & 0 & 0 & 0 & 0 & 0 & 0 & 0 & 0 & 0 & 0 & 0 & 0 & 0 & 0 & 0 & 0 & 0 & 0 & 0 \\
	4 & 0 & 4 & 0 & 0 & 0 & 0 & 0 & 0 & 0 & 0 & 0 & 0 & 0 & 0 & 0 & 0 & 0 & 0 & 0 & 0 & 0 & 0 & 0 \\
	4 & 0 & 0 & 4 & 0 & 0 & 0 & 0 & 0 & 0 & 0 & 0 & 0 & 0 & 0 & 0 & 0 & 0 & 0 & 0 & 0 & 0 & 0 & 0 \\
	4 & 0 & 0 & 0 & 4 & 0 & 0 & 0 & 0 & 0 & 0 & 0 & 0 & 0 & 0 & 0 & 0 & 0 & 0 & 0 & 0 & 0 & 0 & 0 \\
	4 & 0 & 0 & 0 & 0 & 4 & 0 & 0 & 0 & 0 & 0 & 0 & 0 & 0 & 0 & 0 & 0 & 0 & 0 & 0 & 0 & 0 & 0 & 0 \\
	4 & 0 & 0 & 0 & 0 & 0 & 4 & 0 & 0 & 0 & 0 & 0 & 0 & 0 & 0 & 0 & 0 & 0 & 0 & 0 & 0 & 0 & 0 & 0 \\
	2 & 2 & 2 & 2 & 2 & 2 & 2 & 2 & 0 & 0 & 0 & 0 & 0 & 0 & 0 & 0 & 0 & 0 & 0 & 0 & 0 & 0 & 0 & 0 \\
	4 & 0 & 0 & 0 & 0 & 0 & 0 & 0 & 4 & 0 & 0 & 0 & 0 & 0 & 0 & 0 & 0 & 0 & 0 & 0 & 0 & 0 & 0 & 0 \\
	4 & 0 & 0 & 0 & 0 & 0 & 0 & 0 & 0 & 4 & 0 & 0 & 0 & 0 & 0 & 0 & 0 & 0 & 0 & 0 & 0 & 0 & 0 & 0 \\
	4 & 0 & 0 & 0 & 0 & 0 & 0 & 0 & 0 & 0 & 4 & 0 & 0 & 0 & 0 & 0 & 0 & 0 & 0 & 0 & 0 & 0 & 0 & 0 \\
	2 & 2 & 2 & 2 & 0 & 0 & 0 & 0 & 2 & 2 & 2 & 2 & 0 & 0 & 0 & 0 & 0 & 0 & 0 & 0 & 0 & 0 & 0 & 0 \\
	4 & 0 & 0 & 0 & 0 & 0 & 0 & 0 & 0 & 0 & 0 & 0 & 4 & 0 & 0 & 0 & 0 & 0 & 0 & 0 & 0 & 0 & 0 & 0 \\
	2 & 2 & 0 & 0 & 2 & 2 & 0 & 0 & 2 & 2 & 0 & 0 & 2 & 2 & 0 & 0 & 0 & 0 & 0 & 0 & 0 & 0 & 0 & 0 \\
	2 & 0 & 2 & 0 & 2 & 0 & 2 & 0 & 2 & 0 & 2 & 0 & 2 & 0 & 2 & 0 & 0 & 0 & 0 & 0 & 0 & 0 & 0 & 0 \\
	2 & 0 & 0 & 2 & 2 & 0 & 0 & 2 & 2 & 0 & 0 & 2 & 2 & 0 & 0 & 2 & 0 & 0 & 0 & 0 & 0 & 0 & 0 & 0 \\
	4 & 0 & 0 & 0 & 0 & 0 & 0 & 0 & 0 & 0 & 0 & 0 & 0 & 0 & 0 & 0 & 4 & 0 & 0 & 0 & 0 & 0 & 0 & 0 \\
	2 & 0 & 2 & 0 & 2 & 0 & 0 & 2 & 2 & 2 & 0 & 0 & 0 & 0 & 0 & 0 & 2 & 2 & 0 & 0 & 0 & 0 & 0 & 0 \\
	2 & 0 & 0 & 2 & 2 & 2 & 0 & 0 & 2 & 0 & 2 & 0 & 0 & 0 & 0 & 0 & 2 & 0 & 2 & 0 & 0 & 0 & 0 & 0 \\
	2 & 2 & 0 & 0 & 2 & 0 & 2 & 0 & 2 & 0 & 0 & 2 & 0 & 0 & 0 & 0 & 2 & 0 & 0 & 2 & 0 & 0 & 0 & 0 \\
	0 & 2 & 2 & 2 & 2 & 0 & 0 & 0 & 2 & 0 & 0 & 0 & 2 & 0 & 0 & 0 & 2 & 0 & 0 & 0 & 2 & 0 & 0 & 0 \\
	0 & 0 & 0 & 0 & 0 & 0 & 0 & 0 & 2 & 2 & 0 & 0 & 2 & 2 & 0 & 0 & 2 & 2 & 0 & 0 & 2 & 2 & 0 & 0 \\
	0 & 0 & 0 & 0 & 0 & 0 & 0 & 0 & 2 & 0 & 2 & 0 & 2 & 0 & 2 & 0 & 2 & 0 & 2 & 0 & 2 & 0 & 2 & 0 \\
	-3 & 1 & 1 & 1 & 1 & 1 & 1 & 1 & 1 & 1 & 1 & 1 & 1 & 1 & 1 & 1 & 1 & 1 & 1 & 1 & 1 & 1 & 1 & 1 \\
\end{smallmatrix}
\right]
$$

Consider here $\mathfrak{L}_{24,1}^*=4 \mathfrak{L}_{24,1}^{-t}$ and $\mathfrak{L}_{24,2}^*=8 \mathfrak{L}_{24,2}^{-t}$.

We know from the literature that the Leech lattice can be regarded as a sub-lattice of the lattice $\dsE^8 \times \dsE^8 \times \dsE^8$, as in the 8-dimensional case: we use the non-null perturbation the first case ($\scP_{24,1}=\left[\begin{smallmatrix}\scP_{8,1} & 0 & 0 \\ 0 & \scP_{8,1} & 0 \\ 0 & 0 & \scP_{8,1}\end{smallmatrix}\right]$) and the second case to null perturbation ($\scP_{24,2}=\bm O$) and we analyse the performance point of view of the spherical codes, vide Table \ref{table5} (the first two columns refer to the first case).

\begin{table}
$$
\begin{array}{|cc||cc|}
\hline \log_{10} M & \text{distance} & \log_{10} M & \text{distance} \\ \hline 
 10.1917 & 0.633946 & 10.8371 & 0.57735 \\
 15.5128 & 0.484887 & 18.0618 & 0.408248 \\
 19.1994 & 0.370468 & 22.288 & 0.288675 \\
 21.9813 & 0.294144 & 25.2865 & 0.220942 \\
 24.2006 & 0.24225 & 27.6124 & 0.178411 \\
 26.0413 & 0.205264 & 29.5127 & 0.149429 \\
 27.6113 & 0.177774 & 31.1194 & 0.128473 \\
 28.9791 & 0.156625 & 32.5112 & 0.112635 \\
 30.1901 & 0.13989 & 33.7389 & 0.100256 \\
 31.2763 & 0.126336 & 34.8371 & 0.0903175 \\
 32.2609 & 0.115147 & 35.8305 & 0.0821655 \\
 33.161 & 0.10576 & 36.7374 & 0.0753593 \\
 33.99 & 0.097775 & 37.5717 & 0.0695919 \\ \hline 
\end{array}
$$
	\caption{Show spherical code performance 24-dimensional case, for two different representations and respectives perturbations.}
	\label{table5}
\end{table}

\subsection{The lattice $\dsE_6$ and $\dsA_n$}
In the case of $n$-dimensional lattices which do not have full size representation in $n$, for example lattices $\dsA_n$ and $\dsE_6$. The construction will be done by Matrix cholesky decomposition Gram its dual lattice in according Corollary \ref{coroConstruction}. We exemplify this through the lattice $\dsE_6$, note that in this case different perturbations can induce the same number of points but distinct distances, see Table \ref{table6}.

\begin{table}[!h]
$\hfill \begin{array}{|c|c|c|}\hline
w & M & \frac{\delta(\Lambda_w)}{\delta(\dsE_6)} \\ \hline
 1 & 1 & 0.2165 \\
 2 & 16 & 0.3059 \\
 3 & 216 & 0.3761 \\
 4 & 2160 & 0.4011 \\
 5 & 11520 & 0.4538 \\
 6 & 27440 & 0.5469 \\
 7 & 76800 & 0.7639 \\
 8 & 183708 & 0.6381 \\
 9 & 252000 & 0.7376 \\
 10 & 569184 & 0.7247 \\
 11 & 1078272 & 0.6326 \\
 12 & 1514240 & 0.7356 \\
 13 & 2806650 & 0.7436 \\
 14 & 4224000 & 0.7454 \\
 15 & 6714048 & 0.7095 \\
 16 & 9173736 & 0.7707 \\
 17 & 14555520 & 0.8081 \\
 18 & 21294000 & 0.7831 \\
\hline
\end{array}
$
\hfill
$\begin{array}{|c|c|c|}\hline
w & M & \frac{\delta(\Lambda_w)}{\delta(\dsE_6)} \\ \hline
 9. & 252000 & 0.7376 \\
 9.1 & 277200 & 0.6141 \\
 9.2 & 431200 & 0.77 \\
 9.35 & 431200 & 0.7247 \\
 9.4 & 474320 & 0.6437 \\
 9.55 & 474320 & 0.6706 \\
 9.6 & 521752 & 0.6987 \\
 9.7 & 521752 & 0.6643 \\
 10. & 569184 & 0.7247 \\
 10.05 & 569184 & 0.6856 \\
 10.1 & 569184 & 0.6624 \\
 10.3 & 620928 & 0.7601 \\
 10.45 & 698544 & 0.6643 \\
 10.5 & 762048 & 0.7247 \\
 10.65 & 995328 & 0.792 \\
 10.7 & 995328 & 0.7917 \\
 10.85 & 1078272 & 0.6711 \\
 11. & 1078272 & 0.6326 \\ \hline
\end{array} \hfill \,
$
	\caption{Show density rate for 6-dimensional case, for $w$ integer and no integer.}
	\label{table6}
\end{table}

\section{Conclusions}
We conclude that all $n$-dimensional lattice, up less scale, can be approximated by a sequence of lattices that have orthogonal sub-lattice. Furthermore, there is a degree of freedom ($n(n- 1)/2$) for quadratic convergence, this freedom induces quotient group with different number of generators and can make convergency more fast in certain applications, for example in the case of spherical codes are reticulated target has some multiple minimum vectors of some canonical vector, we find a non-null pertubation as it will be more efficient. We present here a method for finding lattices with sub-orthogonal, our method is simpler, more general and more efficient than the one presented in \cite{carina}.



\end{document}